\documentclass[10pt,
twocolumn,
amsmath, 
amssymb, 
superscriptaddress,
aps,
prl,
nofootinbib,
raggedbottom,
longbibliography]{revtex4-2}
\usepackage{graphicx} % Required for inserting images
\usepackage{preamble}
\usepackage{color}
\usepackage[dvipsnames]{xcolor}
\usepackage{cancel}
\usepackage{tikz-cd}
\usepackage[justification=raggedright,singlelinecheck=false]{caption}
\usepackage{ulem}
\usepackage{subfiles}
\usepackage[percent]{overpic}

\usepackage[markup=underlined]{changes}
\definechangesauthor[name=Tin, color=blue]{ts}

\newcommand{\avg}[1]{\left\langle#1\right\rangle}

\begin{document}
\title{Minimal Models of Entropic Order}
\author{Xiaoyang Huang}
\affiliation{Perimeter Institute for Theoretical Physics, Waterloo, Ontario N2L 2Y5, Canada}
\author{Zohar Komargodski}
\affiliation{Simons Center for Geometry and Physics, Stony Brook University, Stony Brook, NY}
\author{Andrew Lucas}
\affiliation{Department of Physics and Center for Theory of Quantum Matter, University of Colorado, Boulder CO 80309 USA}
\author{Fedor K. Popov}
\affiliation{Simons Center for Geometry and Physics, Stony Brook University, Stony Brook, NY}
\author{Tin Sulejmanpasic}
\affiliation{Department of Mathematical Sciences, Durham University, Durham DH1 3LP, UK \looseness=-1}

\begin{abstract}
Due to entropic effects, it is possible that generic high-energy states of a quantum or classical system are ordered.  This leads to spontaneous symmetry breaking at arbitrarily high temperatures. We present minimal models of entropic order that arise from very simple interactions. 
Our main examples are the Arithmetic Ising Model (AIM) and its quantum analogue, where usual Ising spins are replaced by non-negative integers. 
Using a large-flavor expansion together with numerical simulations, we find that the high-temperature phase is ordered in the classical and quantum models. We also introduce classical gas models whose interactions drive the system to a crystal at high temperatures. 
\end{abstract}

\maketitle
\newpage

\section{Introduction}

Ordered phases of matter are commonly assumed to only exist at sufficiently low temperature.
The mechanism is, at face value, obvious: 
at high temperature $T$, the free energy $F=E-TS$ is minimized by maximizing the entropy, and high entropy states are associated with disorder. 

It is possible to prove that the above statements are generally correct, see~\cite{simon2014statistical, dobrushin1968problem} for some rigorous no-go theorems. Central assumptions in these no-go theorems are the finiteness and direct product structure of the space of micro-states at finite volume.

There are physical systems which lie beyond the purview of these theorems. In Quantum Field Theories (QFTs), the Hilbert space is neither factorizable nor finite, allowing order at arbitrarily high temperature \cite{Chai:2020zgq,Chai:2021djc,Chai:2021tpt,Liendo:2022bmv,Ciccone:2023pdk,Komargodski:2024zmt,Hawashin:2024dpp,Hawashin:2025ikp, chaudhuri2021thermal}.\footnote{In the AdS/CFT duality, an ordered high-temperature ensemble maps to a hairy black hole, see for instance~\cite{Buchel:2023zpe}.} In addition, it was recently shown that classical lattice models with an infinite local configuration space can exhibit entropically ordered phases at arbitrarily high temperatures \cite{Han:2025eiw}.  The mechanism that makes this possible is entropic order: by causing one degree of freedom to order, another degree of freedom may fluctuate more strongly.  In this way, the maximal entropy phase, which dominates at high-temperature,  can order. Intermediate-temperature entropic ordering occurs in systems such as helium in the Pomeranchuk effect \cite{pomeranchuk1950theory}, order-by-disorder magnets \cite{villain}, and various colloidal crystals \cite{leo2001}.

Ref. \cite{Han:2025eiw} considered a lattice theory of non-negative integer spins $n_x$ with the classical Hamiltonian $H=\mu \sum\limits_x n_x+U\sum\limits_{\avg{x, y}}n_x^2n_y^2$, and showed that at arbitrary high temperature the square lattice prefers checkerboard-like configurations that have large occupation of order $n_x\sim T$ on one sublattice and $n_x\sim 0$ on the other. 
In the solid phase (the checkerboard phase), there is very little information in which sites are occupied, but there is a lot of information in the occupation number. This allows the solid to have higher entropy than the disordered, gas, phase.

The prospect of realizing entropic order experimentally is tantalizing as it could enable heat- and stress-resistant materials, memory devices and perhaps even high-temperature superconductors \cite{Han:2025eiw}.\footnote{There are also  potential applications in cosmology, see for instance~\cite{Meade:2018saz, Carena:2021onl}.}  Motivated by this possibility, we 
present minimal models of entropic order, relying on simple ingredients that may be realizable in experiments (such as Rydberg atom systems where $n_x$ is the level of an atom located at position $x$).
Firstly 
we restrict ourselves to repulsive  quadratic interactions such as $\sum\limits_{\small  \avg{x, y}} n_xn_y$, which are more natural in models of physical significance (e.g. extended Bose-Hubbard models, Rydberg atom arrays, etc.).
Since $n_x$ could potentially represent a boson occupation number or the excitation level of an atom, such quadratic interactions should be more realistic.
Next, we go beyond classical statistical systems and consider analogous quantum models, where we show that the same phenomena take place. Finally, we discuss classical gas models with two-body interactions
showing that they, too, can be in a solid state at arbitrarily high temperature.
The quantum models and the classical gas models show that entropic order is  robust in the presence of Hamiltonian dynamics.

\section{The Arithmetic Ising Model}

Consider an Ising-type model on a 2D square lattice\footnote{
We expect the conclusions to qualitatively extend to other lattices, but the universality class of the transition is possibly sensitive to the microscopic lattice, as in  the theory of hardcore lattice gases \cite{baxter1982}.}
\begin{equation}
    H(n_x) := \mu\sum_x n_x + U\sum_{\langle x, y\rangle}n_x n_y \label{eq:Hn+nn}
\end{equation}
where $U,\mu\ge 0$ are positive energy constants and $n_x\in \lbrace0,1,\ldots\rbrace $ are non-negative integers. The symbol $\langle x, y\rangle $ indicates that lattice sites $x$ and $y$ are next-neighbors.  We call this model the Arithmetic Ising Model (AIM).   

\begin{figure}
    \centering
    \includegraphics[width=\linewidth]{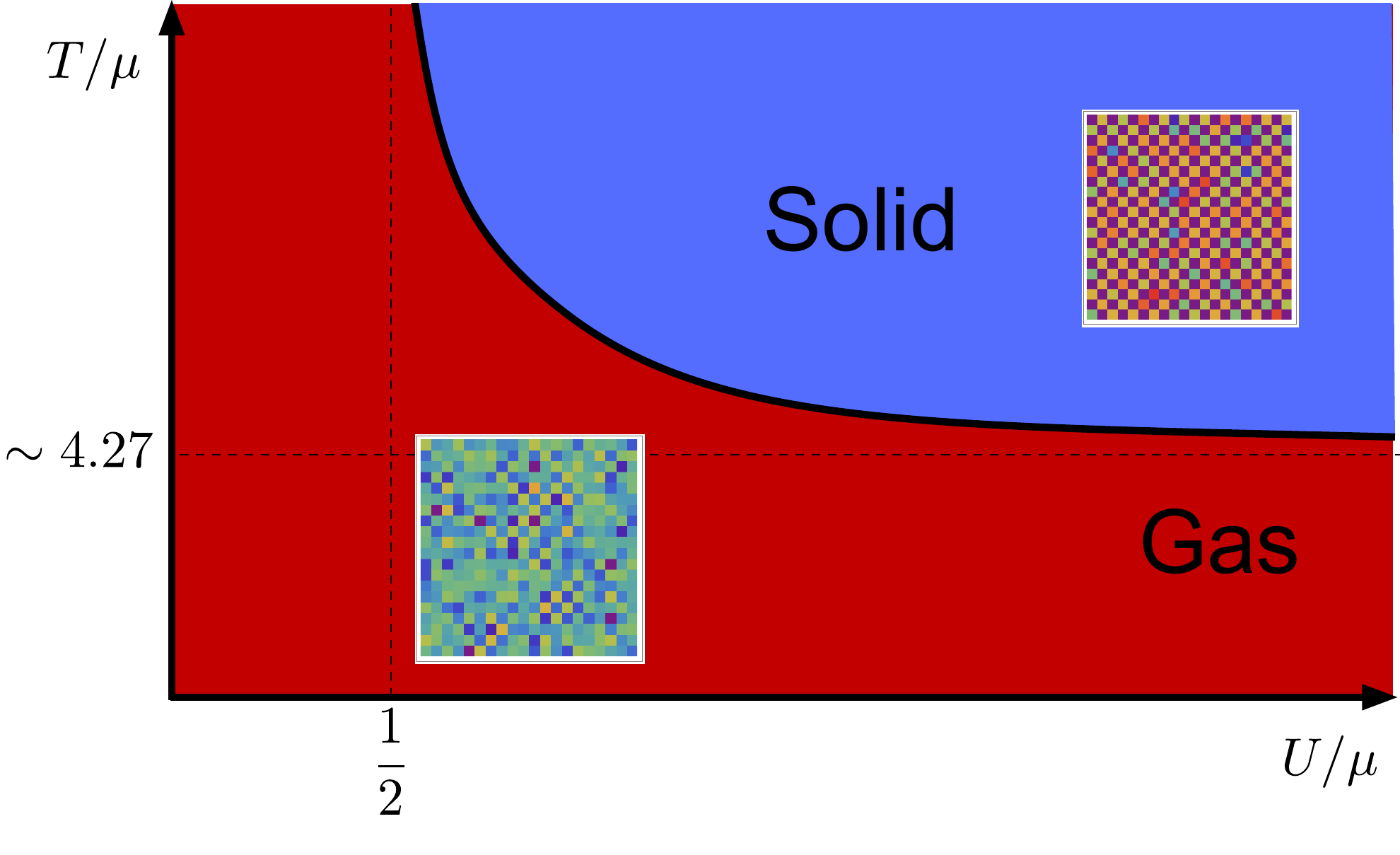}
    \caption{A schematic phase diagram of the model~\eqref{eq:Hn+nn}. The insets show typical Monte Carlo configurations of the two phases. }
    \label{fig:phase_diag}
\end{figure}

We are interested in sampling from the thermal (Gibbs) ensemble at inverse temperature $\beta$: \begin{equation}
    \mathbb{P}_\beta(n_x) := \frac{1}{Z(\beta)}\mathrm{e}^{-\beta H(n_x)}
\end{equation}
where the partition function \begin{equation}
    Z(\beta) := \mathrm{e}^{-\beta F(\beta)} = \sum_{\{n_x\}}\mathrm{e}^{-\beta H(n_x)}.
\end{equation}

Although the state space of all possible $\{n_x\}$ at finite volume is countably infinite, it is easy to see that the above problem is well-posed for any $\beta>0$.\footnote{The partition function is a finite number for any finite $L$ because $Z(\beta) \le Z(\beta)|_{U=0} = (1-\mathrm{e}^{-\beta \mu})^{-L^2}$.} 

This model exhibits an unconventional phase diagram. To begin with, it is straightforward to see that the unique ground state of the AIM corresponds to the empty configuration, $n_x = 0$, for all sites.  Consequently, at low temperatures, the system behaves as a disordered lattice gas. As the temperature increases, sites on the lattice become increasingly occupied by nonzero $n_x$. However, the presence of the repulsive $U$-term energetically penalizes the simultaneous occupation of neighboring sites.  When\footnote{We  explain the origin of this estimate below.}  $U/\mu\gtrsim 
\frac12$ the model favors a high-temperature phase in which the system organizes into a ``checkerboard'' solid: sites on one sublattice host a large occupation number, $\langle n_x \rangle \sim T$, while the sites on the complementary sublattice are nearly empty. This solid phase spontaneously breaks the translation symmetry of the square lattice and, as such, it constitutes an ordered phase that persists at high temperatures.

To see this, we study the system in Mean-Field Theory (MFT) approximation. Anticipating the solid phase checkerboard phase, we construct two mean fields $n_x\approx \bar n_{A,B}$, when $x$ is on sublattice $A$ and $B$ respectively. The energy density of such a MFT is given by
\begin{equation}
    e= 2 U n_A n_B+\frac{\mu}{2}(n_A+n_B)
\end{equation}

\begin{figure}
    \centering
    \includegraphics[width=1.0\linewidth]{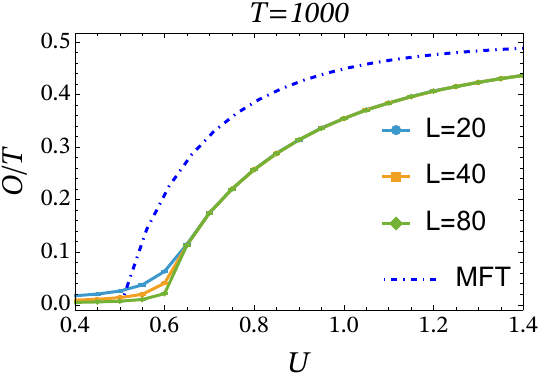}
    \includegraphics[width=1.0\linewidth]{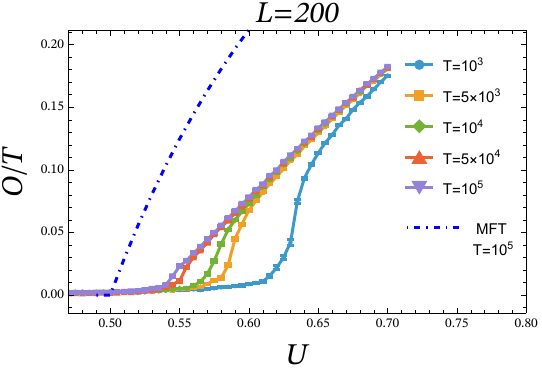}

    \caption{Top: The MC measurements of the order parameter as function of $U$ at $T=1000$ (given in units of $\mu$), for lattice sizes $L=20,40,60$ and $80$. The results are compared to a MFT prediction (see SM for details). Bottom: Monte Carlo comparison of transitions at different temperatures, showing a slow drift towards $U/\mu=1/2$.}
    
\label{fig:order_par_T=1000}
\end{figure}

To find the entropy of the mean field configuration, we note that the number of ways we distribute  $n_{A,B}L^2/2$ integers into $L^2/2$ sites, given by
\begin{equation}
     e^{S_{A,B}}\approx e^{(n_{A,B}+1)\log(n_{A,B}+1)-n_{A,B}\log n_{A,B}}
\end{equation}
where we took the large $L^2$ limit. The entropy density is given by $s=(S_A+S_B)/L^2$. The free energy density is $f=e-Ts$ can then be minimized w.r.t. $n_A$ and $n_B$. Careful analysis shows that at large $T$, the minimum is achieved at $O=n_A-n_B\ne 0$ as long as 
% Let us give a mean-field theory (MFT) argument for the solid phase, leaving technical details to the Supplementary Materials (SM).
% First we estimate the contribution of the gas phase to the partition function; in this regime, $\langle n_x\rangle =\bar n$ is $x$-independent. MFT predicts that at high temperature \begin{equation}
%     \bar n = \sqrt{T/U}/2, \label{eq:barn}
% \end{equation}
% \ts{changed to correct factor}
% so 
% \begin{equation}\label{gasMFT}
%     \log Z^{\rm MFT}_{\text{gas}}\approx \frac12L^2 \log\left(\frac{T}{4U}\right)~.
% \end{equation}
% \ts{changed the factor correctly and put propto}
% Now let us check the contribution of the solid phase.  In MFT, $\langle n_x\rangle\sim \bar n_{A,B}$ depending on whether the site $x$ is on the A or B sublattice, and we find that $\bar n_A\sim T/\mu $ and $\bar n_B \sim 1$.  Therefore, the partition function at high temperature is given by
% \begin{equation}\label{solidMFT}
%     \log Z^{\rm MFT}_{\mathrm{solid}}\propto\frac12L^2 \log\left(\frac{T}{\mu}\right)~.
% \end{equation}
% \ts{Clearly if $U$ is sufficiently large the solid phase will dominate the partition function. Careful MFT analysis (SM) shows this happens when}
\begin{equation}\label{eq:Umu_inequality}
    U/\mu>1/2~.
\end{equation}
$U\ge \mu/2$ (see dashed line in Fig.~\ref{fig:order_par_T=1000}).
Thus, MFT predicts that as long as~\eqref{eq:Umu_inequality} is satisfied, the high-temperature phase is a solid.

When $U\rightarrow\infty$ the model is related to an exactly solvable hardcore lattice gas on the square lattice \cite{baxter1980hard,baxter1982}. In this limit the solid occurs for $T\gtrsim T_c\approx 4.27\mu$.

In summary, the phase diagram of the model as a function of $T/\mu$ and $U/\mu$ is shown in Figure~\ref{fig:phase_diag}.
A detailed analysis reveals that the phase transition line is second-order everywhere and in the Ising universality class.  The transition is driven by fluctuations in $n$ that add particles on the A/B sublattice and subtract them on the B/A sublattice.

Surprisingly, as we will see below, the MFT estimate~\eqref{eq:Umu_inequality} appears to be exact in the  $T\rightarrow\infty$ limit.   
To justify that claim, let us turn to a variant of AIM: the ``colored AIM"   model with $k$ species of particle on every site: we let $n_x= \sum\limits_{\alpha=1}^kn_{x,\alpha}$, where $n_{x,\alpha}$ are  nonnegative integers. In addition we rescale $U\to U/k$.
The Hamiltonian becomes
\begin{gather}
    H_k =\mu \sum_{x,\alpha} n_{x,\alpha} + \frac{U}{k} \sum_{\langle x,y\rangle,\alpha,\beta}  n_{x,\alpha} n_{y,\beta}~,
\end{gather}
and we find that
\begin{align}
    Z _k &= \sum_{\left\{n_{x,\alpha} \right\}} e^{-\beta H_k(n_{x,\alpha} )} \notag \\
    &= \sum_{\left\{n_{x} \right\}} e^{-\beta H_1(n_{x})} \prod_i\binom{n_x + k - 1}{ k -1}~.
\end{align}
Note that the model at $k=1$
coincides with the original AIM. The additional species modify the microstates counting, leading to simplifications as $k\to\infty$.
Thus, let us now show that MFT becomes exact as $k\to \infty$.
To see that we rescale the density as $n_i = k \rho_i$ and replace the sum by an integral over $\rho_x$.  As $k\rightarrow\infty$ we obtain 
\begin{gather}\begin{split}
    & Z \approx \int \prod d\rho_i \exp\biggl(- k \biggl(\beta\mu\sum_x \rho_x + \beta  U \sum_{\langle x, y\rangle} \rho_x \rho_y +  \\ &\sum_i\left(\rho_x + 1\right) \log(\rho_x + 1) -  \sum_x\rho_x \log \rho_x \biggr)\biggr)~.\end{split} \label{eq:EFTlargek}
\end{gather}
Therefore $k$ acts as a loop expansion parameter and at $k\to \infty$ we just need to minimize the ``action.'' The minimization gives $\bar\rho=\sqrt{T/4U}$ at hight $T$ while fluctuations are suppressed at large $k$ allowing systematic study. Indeed, detailed computation shows that the solid phase happens when
\begin{equation}
\frac{U}{\mu} > \frac{1}{2} + \frac{\sqrt{\beta \mu}}{2\sqrt{2}\pi k } \log\frac{1}{\beta\mu}    + \cdots, \label{eq:Umu_improved}
\end{equation}
in the $1/k$ expansion.
We see that to the first nontrivial order in the $1/k$ expansion, the MFT result~\eqref{eq:Umu_inequality} remains valid at $T\to\infty$.

The numerical results below are qualitatively consistent with \eqref{eq:Umu_improved} even at $k=1$. Moreover, in the large-$k$ limit the theory appears to develop an exact continuous symmetry as $T \to \infty$. This symmetry gives rise to a parametrically light mode in the high-temperature gas phase, rendering it qualitatively distinct from a conventional high temperature gas, and appears to protect the infinite-temperature transition at $U=\mu/2$ from perturbative  $1/k$ corrections. We briefly comment on this structure in the SM and defer a detailed analysis of its origin and implications to future work.

\begin{figure}

    \includegraphics[width=1.0\linewidth]{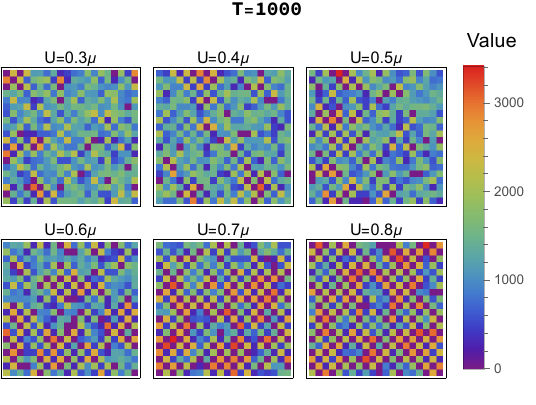}
    \caption{Typical Monte Carlo configurations of the AIM at various values of $U$ (with $\mu=1$). }
\label{fig:configs_T=1000}
\end{figure}

Let us now turn to the numerical results.
We use Monte Carlo simulations to numerically draw from the thermal ensemble of AIM directly at $k=1$.  Extensive simulations confirm that there is high-temperature order when $U/\mu$ is sufficiently large: see Fig.~\ref{fig:configs_T=1000}.  More quantitatively, the transition to a solid is further confirmed by the measurements of the order parameter $O=\frac{1}{2L^2}\left(\sum_{x\in A} n_x-\sum_{x\in B} n_x\right)$, where $A$ and $B$ are the two sublattices. We plot the order parameter as a function of $U/\mu$ at different values of temperature in Fig. \ref{fig:order_par_T=1000}. The numerical results are qualitatively consistent with \eqref{eq:Umu_improved}, even for $k=1$. 

\begin{figure}[t]
    \centering
    \includegraphics[width=\columnwidth]{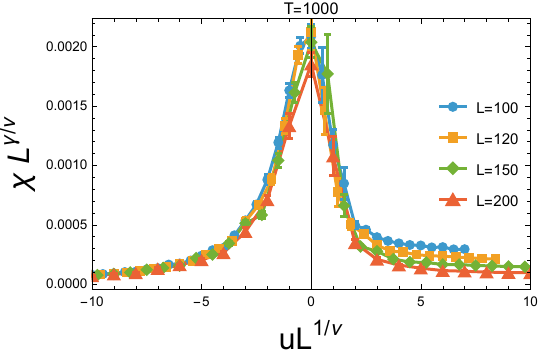}
    \caption{Order parameter susceptibility scaling for the 2D AIM with 2D Ising critical exponens $\nu=1,\gamma=7/4$, and $U_c=0.63$. }
    \label{fig:chi}
\end{figure}

Further we look at the order parameter susceptibility, defined by $\chi=L^2(\avg O^2-\avg{O}^2)$. The susceptibility $\chi$ obeys the well known finite size scaling near the critical point
\[
\chi=L^{\gamma/\nu}F(u L^{1/\nu})
\]
where $u=(U-U_c)/U_{c}$ is the reduced coupling driving the transition. In Fig.~\ref{fig:chi} we show a collapse of data for multiple volumes is achieved when 2D Ising critical exponents are used.

One could ask what if $n_x$ in the Arithmetic Ising Model~\eqref{eq:Hn+nn} is bounded by some $0\leq n_x\leq N$. In this case the disordered phase with uniform density would eventually dominate, consistent with the aforementioned no-go theorem \cite{dobrushin1968problem, simon2014statistical} on high-temperature order. We can estimate in MFT the critical temperature when $N$ is large. We find that the solid is expected to exist in the window
$\mu\lesssim T\lesssim NU$ (as long as $U/\mu \gtrsim 1$).
This follows from the fact that the solid's entropy cannot continue growing past $\frac{1}{2}L^2 \log N$. 

Another perturbation that eventually destroys the entropically-ordered solid is to add an on-site quadratic repulsion energy: \begin{equation}
H \rightarrow H + K \sum_x n_x^2.    
\end{equation}
This perturbation changes the high-temperature free energy of the solid to $\log Z^{\rm MFT}_{\text{solid}}\sim \frac{1}{4}L^2 \log \frac{T}{K}$. Therefore, the disordered gas state always dominates at high enough temperature. We can estimate that the solid melts at $T_c \sim \frac{U^2}{ K}$.  For sufficiently small $K$ the melting temperature is arbitrarily high. Similarly, small next-to-next to nearest neighbor interactions are not dangerous if they are sufficiently small.

\subsection{Quantum Arithmetic Ising Model (qAIM)} 
The entropically-ordered solid is robust in the presence of single-particle quantum tunneling between adjacent lattice sites.  Explicitly, we consider a quantum boson model on the 2d square lattice 
\begin{align}
    H& = H_0 - t V\nonumber-\kappa V', \quad 
    H_0  = \mu\sum_x n_x + U \sum_{\langle x, y\rangle}n_x n_y \nonumber\\
    V& = \sum_{\langle x , y\rangle}\left(b^\dagger_x b_y+\mathrm{h.c.}\right), \quad   V' = \sum_i \left(b_x+b_x^\dagger\right),
\end{align}
with $t$ the hopping strength. The term $V'$ is  introduced to break the $U(1)$ symmetry.  As before $\mu,U>0$.

The zero temperature ground state, at $\kappa=0$ and $\mu>2t$, has no particles at all. As the temperature increases an increasing number of particles appear in the system and we can ask whether, when \eqref{eq:Umu_inequality} is obeyed, a solid again forms at high temperature in the quantum model.  Indeed, % we find that this is the case: e.g. 
we can estimate $ b^\dagger_i b_j\sim \sqrt{\bar n_A \bar n_B} $ suggesting that at high temperatures where $\sqrt{\bar n_A \bar n_B}\sim \sqrt{T}$ (in both the solid and gas phases), the quantum fluctuations become increasingly negligible compared to all terms in $H_0$. Similarly, the term $V'$ is also suppressed at high temperature. Additionally, a $k$-colored generalization of qAIM with the following Hamiltonian

\begin{multline}
H = -t\sum_{\langle x,y\rangle,\alpha}\!\!\big(b^\dagger_{x,\alpha} b_{y,\alpha}+{\rm h.c.}\big)\\
      +\mu \sum_{x,\alpha} n_{x,\alpha}
      +\frac{U}{k}\sum_{\langle x,y\rangle}\sum_{\alpha,\beta} n_{x,\alpha} n_{x,\beta}.
\end{multline}
where $\alpha=1,\cdots, k$ and $n_{i,\alpha}=b^\dagger_{x,\alpha}b_{x,\alpha}$ can be treated in the large $k$ approximation (see SM) and can be shown to have a solid phase at high $T$ as long as \eqref{eq:Umu_inequality} is obeyed.\footnote{We can also add the term $V'$ to the large $k$ analysis without changing the conclusions.}

\section{Continuous gas models}

\subsection{A Gas of "Polymers"}

Now we turn to a classical gas model which we will see  has a high-temperature entropic order. 

We consider a gas of extended objects, which we dub polymers. The internal size of the extended object is denoted by $\rho_i$ and the center of mass position is denoted by $x_i$ with $i=1,...,N$. 

We consider 2-body interactions: $V_{ij}=\rho_i^2 \rho_j^2 v(|\bm x_i-\bm x_j|)$. For instance, we can take $v(r)=U_0\Theta(R-r)$, where $R>0$ and $\Theta(x)$ is the Heaviside step function. This means that the folded polymers  repel only when their centers come within distance $R$ of each other. (This is of course a very crude model, but the conclusions are robust to many deformations thereof.) Crucially the strength of the repulsion grows as $\rho_i^2\rho_j^2$.

We assume that folded polymers in isolation tend to shrink with energy that goes as ${\kappa\over 2}\rho_i^2$.

The canonical partition function, up to an unimportant constant, is given by
\begin{multline}\label{eq:Z_class_spring}
    Z= \prod_{i=1}^N\int d^3\bm x_i\int_0^\infty d\rho_i  \\\times\exp\left(-\sum_{\substack{i=1\\i>j}}^N\rho_i^2\rho_j^2 v(r_{ij})\beta-\sum_{i=1}^N \frac{\kappa}{2}\rho_i^2 \beta\right)\;,
\end{multline}
To analyze this model, note that non-overlapping particles have a weight $w_1=\int d\rho e^{-\frac{\kappa}{2}\rho^2\beta}\sim T^{1/2}$. If two particles overlap, the weight is $w_2=\int d\rho_1d\rho_2  e^{-\rho_1^2\rho_2^2\beta-\frac{\kappa}{2}(\rho_1^2+\rho_2^2)\beta}\sim T^{\frac{1}{2}}\log(T)$. It is clear that $w_2/w_1^2\rightarrow 0$ when $T\rightarrow \infty$, so that overlaps are completely suppressed, and the model effectively becomes a hard sphere model, which is known to exhibit a solid phase in 3D at sufficiently large packing ratio \cite{alder1957phase,wood1957preliminary,likos2001effective}. In our model, therefore, for sufficiently large densities of the polymers, we will obtain a solid state at high temperature.

If the function $v(r)$ is allowed to be any short-ranged repulsive potential, the two-particle weight has nontrivial dependence on the particle's relative distance $r$, and behaves as  $w_2(r)\sim T$ when $\kappa^2 \beta\gg v(r)$ (i.e. long distance) and $w_2(r)\sim T^{1/2}\log T$ when $\kappa^2 \beta\ll v(r)$ (short distance). This is exactly the same behavior as for step function, except now we have an effective radius given by  $v(R_{\rm eff})\sim \kappa^2\beta=\kappa^2/T$. Because $v(r)$ is a monotonically decreasing function, we have that $R_{\rm eff}$ grows with temperature, effectively increasing the packing ratio for any density. Hence in such a system we expect high temeprature order to emerge at any density, as long as the temperature is sufficiently high.

\subsection{Grand Canonical Ensembles}

Consider a classical grand canonical partition function of indistinguishable particles with a fixed chemical potential
\begin{equation}
\sum_N {z^N\over N!} \int d^{3N}r \, \exp\left(-\beta \sum_{i<j}U\left(\left|\vec r_i-\vec r_j\right|\right)\right)
\end{equation}
where $z=e^{\mu/T}
\left({mT\over2\pi\hbar^2}\right)^{3/2}$, $\mu$ is the chemical potential,\footnote{We note that it may be difficult to maintain a constant chemical potential in realistic situations. 
} $m$ is the particle mass and $\hbar $ is  Planck's constant.

Using a field theory identity (see SM) we can employ MFT and look for minima of the free energy
\begin{gather*}
    \mathcal{ F} =\frac12 \iint d^3 x d^3 y\, n(x)n(y)U(|\vec x-\vec y|)+ \notag \\ 
    T\int d^3 x  n(x)(\log(n(x)/z)-1)~.
\end{gather*}
where $n$ is the density of particles. Assuming uniform $n$ and high temperature, we can find a saddle point $\bar n={T\over \mathcal U}\log(z\mathcal U/T)$ where $\mathcal U=\int d^3\vec x U(|\vec x|)$. This gives the free energy 
\begin{equation}\label{eq:Fgas}
    \mathcal F_{0}=\frac{1}{2}\bar n^2\mathcal UL^3+T\bar n\log(\bar n/z)L^3\sim -T^2(\log T)^2L^3
\end{equation}

Now let us study fluctuations on top of the uniform saddle  $n(x)=\bar n+\delta n(x)$. To quadratic order the free energy is
\begin{gather*}
\mathcal F=\mathcal F_0+\iint d^3 x\, d^3 y\, \delta n(\vec x)\,U(|\vec x-\vec y|)\,\delta n(\vec y)+\notag\\+T\int  d^3 x\frac{\delta n(x)^2}{\bar n}
\end{gather*}
Now since $\bar n\sim T\log T$ at high temperature, the second term is sub-leading. Therefore the uniform phase is destabilized if there exist modes $\delta n(\vec x)$ for which the first term above is negative. This precisely happens when $U(|\vec r|)$ has negative Fourier coefficients. This occurs, among others, for a step function potential $U(r)=U_0\Theta(R-r)$. The unstable mode will drive the system to a cluster solid at sufficiently high temperature.

What happens physically is that the density increases as we increase the temperature, and the uniform gas becomes unstable to forming a solid due to the repulsion as in the classical theory of freezing~\cite{alexander1978should}, except that here it happens at high temperature due to a constant chemical potential. That the melting temperature in such models increases with density was also discussed in \cite{van2009cluster}.

\section{Conclusions} 
We showed multiple minimal models for entropic order and demonstrated the robustness of the effect to classical and quantum kinetics.  A promising quantum realization of these models may be Rydberg atoms trapped in optical lattices, where $n_x$ is the principal quantum number of the atom.  However, unlike in previous uses of Rydberg blockade to realize exotic spin liquids \cite{semeghini2021probing,Giudici:2022olb}, here it will be important to ensure a repulsive Rydberg blockade for many distinct Rydberg states \cite{Lahaye:2016ccd,rydberg2} in order to realize the high-temperature quantum solid.  Alternatively, in continuous space the principal quantum number may play the role of $\rho_x$ in the otherwise classical ``polymer gas".

\section*{Acknowledgements}

We thank  E. Andriolo, L. Homeier, N. Nekrasov, M. Nguyen and E. Richards for discussions. 
AL is supported by the National Science Foundation under CAREER Grant DMR-2145544. TS is supported by Royal Society University Research Fellowship and STFC consolidated grant number ST/X000591/1. Research at Perimeter Institute is supported in part by the
Government of Canada through the Department of Innovation,
Science and Economic Development and by the Province of Ontario
through the Ministry of Colleges and Universities.
\bibliography{refs.bib}

\onecolumngrid
\newpage
\begin{appendix}

\begin{center}
   \Large \bf SUPPLEMENTAL MATERIALS
\end{center} 

%!TEX root = PRL-v2.tex
\section{The Colored AIM and the large $k$ expansion}
Let us consider the case of the large number of colors. To do that we will assume that we have $k$ particle species per each site. Then we introduce the following energy functional
\begin{gather}
    E(n_{x,\alpha}) = \mu \sum_{x,\alpha} n_{x,\alpha} + \frac{U}{k} \sum_{\avg{x, y},\alpha,\beta}  n_{x,\alpha} n_{y,\beta},
\end{gather}
where $x$ is the site multiindex and $\alpha$ is the species index. %Note that without the loss of generality we absorbed $\mu$ into the definition of temperature. 
Then we first introduce the $n_x =\sum\limits_\alpha n_{x,\alpha}$. At first glance, it does not change the resulted energy functional. What actually changes is the entropy factor. For instance if $k=2$ we have now $n_x = n$ particles on one site, it corresponds to $n+1$ configurations instead of just $1$. For general $k$ this degeneracy is $\binom{n+k-1}{k-1}$. Thus we get that our partition function has the following form,
\begin{gather}
    Z = \sum_{\left\{n_{i,\alpha} \right\}} e^{-\beta E(n_{i,\alpha})} = \sum_{\left\{n_{i} \right\}} e^{-\beta H(n_{i})} \binom{n_x + k - 1}{ k -1},
\end{gather}
we see that effectively the Hamiltonian of k-AIM model is essentially the same as a Hamiltonian of AIM model but with different distribution of states. Thus where the binomial coefficient takes into account the different number of species. Not let us label $n_x = k \rho_x$, so that then we can replace the sum over $n_x$ by an integral $\rho_x$\footnote{That should only produce some exponentially suppressed corrections to that approximation. } and apply Stirling approximation for the factorials. It gives as the following approximation
\begin{gather}
    Z \approx \int \prod d\rho_x \exp\left(- k \left(\beta\sum_x \rho_x + \beta  U \sum_{\avg{x,y}} \rho_x \rho_y - \big(  \sum_x\left(\rho_x + 1\right) \log(\rho_x + 1) -  \sum_x\rho_x \log \rho_x \big) \right)\right)
\end{gather}

If we assume the gas phase, and set $\rho_x=\rho$, the saddle point equation is given by
\begin{equation}
    \rho = \frac{1}{e^{\beta+4\beta U \rho} - 1}
\end{equation}
which can be solved order by order in small $\beta$
\begin{equation}
    \rho = \frac{1}{2\sqrt{\beta U}} - \frac{\mu+2 U}{8U} + \frac{28 U^2-12 U \mu + 3\mu^2}{192 U^{3/2}} \sqrt{\beta} + \mathcal{O}(\beta^\frac32)\;.
\end{equation}
Now we want to study the fluctuations on top of this uniform state. For that we introduce $\rho_x = \rho + \frac{1}{\sqrt{k \beta U}}\phi_x$. That gives that the effective Hamiltonian is
\begin{gather}
   H(\phi_x) \approx    \sum_{i\sim j}\phi_x \phi_y + \sum_x \Big\{\frac{ \phi _x^2}{2 \rho (1+\rho) \beta U} + \underbrace{\frac{1}{6 \sqrt{k}(\beta U)^\frac32}
   \left(\frac{1}{(\rho +1)^2}-\frac{1}{\rho ^2}\right)}_{\mu_3} \phi _x^3 +  \underbrace{\frac{1}{12 k(\beta U)^2}\left(\frac{1}{\rho ^3}-\frac{1}{(\rho +1)^3}\right)}_{\lambda_4} \phi _x^4 \Big\}
\end{gather}
Diagonalizing this Hamiltonian in the Fourier space yields that we have the following propagator
\begin{gather}
    G^{-1}_0(k) = 4 + 2 \cos k_x + 2 \cos k_y + m_0^2, \notag\\
    m^2_0 = \frac{1}{\beta U\rho(1+\rho)} - 4 = 2(\mu-2 U) \sqrt{\frac{\beta }{U}}+ \frac{\beta(3 \mu^2 - 12 \mu U  + 28U^2)}{6U}+\cO(\beta^{\frac{3}{2}}) \label{eq:prop}
\end{gather}
and the minimum occurs around the point $k_\pi = (\pi, \pi)$. From this we see that in the large $k$ limit the phase transition should happen at the point
\begin{align}\label{eq:msq0_corr}
    m^2_0 = 0  \quad U^{(0)}_c = \frac12 + \frac{\sqrt{\beta \mu^3}}{3\sqrt{2}} +\cO(\beta)
\end{align}
Now, let us take into account the corrections in the leading $\frac{1}{k}$ regime. The leading corrections of the order $\frac{1}{\sqrt{k}}$ that come from $\phi^3$ interaction vanish. Then the next leading diagrams, that contribute to the free energy, have the following form:
\begin{gather}
    \Sigma(p) = \Sigma_a(p) + \Sigma_b(p) + \Sigma_c(p)  = \notag\\
    \begin{tikzpicture}
    \draw (-1,0) -- (1,0);         
    \draw (0,0) -- (0,1);          
    \draw (0,1.5) circle (0.5);         
\end{tikzpicture} \quad + \quad \begin{tikzpicture}
    \draw (-1.5,0) -- (-0.5,0);
    \draw (0.5,0) -- (1.5,0);
    \draw (0,0) circle (0.5);
\end{tikzpicture} \quad + \quad  \begin{tikzpicture}
    \draw (-1,0) -- (1,0);          
    \draw (0,0.5) circle (0.5);         
\end{tikzpicture}
\end{gather}
We can compute these diagrams at the point $p = k_\pi$ and get 
\begin{gather}
    \Sigma_a(p=k_\pi) = \frac{1}{2} (-6\mu_3)^2 G_0(0) \int \frac{d^2 k}{(2\pi)^2} G_0(k) = \frac{18 \mu_3^2}{8 + m_0^2} I(m_0^2)  , \notag\\
    \Sigma_b(p=k_\pi) = \frac{1}{2}(-6\mu_3) ^2 \int \frac{d^2 k}{(2\pi)^2} G_0(k) G_0(k+k_\pi) = \frac{18 \mu_3^2}{4+m_0^2} I(m_0^2) , \notag\\
    \Sigma_c(p=k_\pi) = \frac12 (-24\lambda_4) \int \frac{d^2 k}{(2\pi)^2} G_0(k) = -12 \lambda_4  I(m_0^2)
\end{gather}
where
\begin{gather}
I(m_0^2) = \int \frac{d^2 k}{(2\pi)^2} \frac{1}{4 + 2 \cos k_x + 2 \cos k_y + m_0^2} =
\frac{2 K\left(\frac{16}{\left(m_0^2+4\right)^2}\right)}{\pi  \left(m_0^2+4\right)} = \frac{1}{4\pi}  \log \frac{32}{m_0^2} + \mathcal{O}(m_0^2).
\end{gather}
For that we notice that
\begin{gather}
    \Sigma(k_\pi) = 6 I(m_0^2)\left(\frac{3\mu_3^2}{8 + m_0^2} + \frac{3\mu_3^2}{4+m_0^2} -2 \lambda_4 \right) = - \frac{8 \mu \beta}{k} I(m_0^2) + \mathcal{O}(\beta^2)\;,
\end{gather}
The most important part here is that the self-energy $\Sigma(k_\pi) \propto \beta$ as $\beta\to 0$, thus the correction to the critical mass will vanish in the limit $\beta \to 0$ and we expect that $U^c_\infty = 0.5$. Effectively in the leading $\frac{1}{k}$ expansion we find the following field theory in 2 dimensions
\begin{gather}
    S = \frac12 (\partial_\mu \phi)^2 + \frac12 m_0^2 \phi^2 + \frac{\lambda}{4!}\phi^4, \quad  \lambda = \frac{16 \beta \mu }{k}, \quad m_0^2 =  2(\mu-2 U) \sqrt{\frac{\beta }{U}}+ \frac{\beta(3 \mu^2 - 12 \mu U + 28U^2)}{6U} \label{eq:bare_action}
\end{gather}
Then we can use the results of \cite{RychkovVitale2016,PelissettoVicari2015} to find the critical value of $m_0^2$. Let us briefly explain how the computations of the crtical mass is performed, for that we first renormalize our theory \eqref{eq:bare_action} by introducing the normal ordered operators
\begin{gather}
    S = \frac12 (\partial_\mu \phi)^2 + \frac12 m^2 :\phi^2:_m + \frac{\lambda}{4!}  :\phi^4:_m, \quad
    m^2 = m_0^2 + \frac{\lambda}{8\pi} \log \frac{32}{m^2}
\end{gather}
The normal ordered operators mean that when we study the Feynman diagrams, we omit all diagrams with a tadpole insertion. Then the first diagram in this case comes from
\begin{align}
    &\Sigma(p = 0) =  \begin{tikzpicture}[baseline={(0,-0.1)}]
    \draw (-1,0) -- (1,0);          
    \draw (0,0) circle (0.5);         
\end{tikzpicture} = \frac{\lambda^2}{6} \int \frac{d^2 p}{(2\pi)^2} \frac{d^2 q}{(2\pi)^2} \frac{1}{p^2 + m^2}\frac{1}{q^2 + m^2} \frac{1}{(p+q)^2 + m^2} = C \frac{\lambda^2}{m^2},\nonumber \\ &C = -\frac16 \int\limits^1_0 d x~ \frac{\log(x(1-x))}{16\pi^2(1  + x^2 - x)} \approx 0.0025
\end{align}
That gives the condition for the critical mass as
\begin{gather} 
     G^{-1}(p = 0) = G_{m^2}^{-1}(p=0) - \Sigma(p=0) = m^2 - C \frac{\lambda^2}{m^2} = 0,\quad 
    m^2_{\rm crit}  = \sqrt{C} \lambda
\end{gather}
Then pluggin back to the orginal theory we get
\begin{gather}
    m^2_{\rm crit,0} = \sqrt{C}\lambda - \frac{\lambda}{8\pi} \log \frac{32}{\sqrt{C}\lambda}  = \frac{\lambda}{8\pi} \log \frac{\sqrt{C}\lambda e^{8\pi \sqrt{C}}}{32}
\end{gather}
Plugging the value of the effective coupling constant and how does $m_{0}^2$ depends on $U$  we find
\begin{gather}
    m_{\rm crit,0}^2(\beta)
    = \frac{2\beta}{\pi k} \log \frac{\sqrt{C} e^{8\pi\sqrt{C}} \beta \mu}{2k}
\end{gather}
where we have used $k=1$ and \eqref{eq:prop}. The large temperature expansion of this critical mass is
\begin{gather}
    U = \frac{\mu}{2} + \frac{\sqrt{\beta \mu^3}}{2\sqrt{2} \pi k} \log\frac{1}{\beta \mu}, \quad \beta \gg 1
\end{gather}

\begin{figure}
    \centering
    \includegraphics[width=0.5\linewidth]{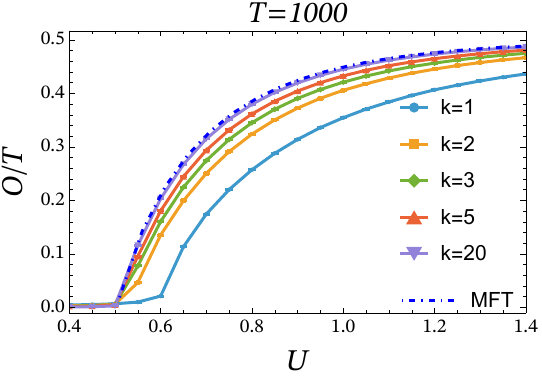}
    \caption{Monte Carlo simulation at system size $L=80$ results for the order parameter, for multiple values of $k$. All dimensionful quantities are in units of $\mu$. }
    \label{fig:MC_large_k}
\end{figure}

Finally, let us present the Monte Carlo results for various values of $k$. In Fig.~\ref{fig:MC_large_k} we show Monte Carlo simulation results for the order parameter at multiple values of $k$, along with the MFT prediction. We see a rapid approach of the large color results to the MFT prediction.

Before concluding this section we note that the robustness of the transition at $U=\mu/2$ when $\beta\rightarrow 0$ seems to be a result of a curious continuous symmetry in this limit. Indeed, there seems to exist an exact continuous symmetry $\phi_x\rightarrow (-1)^{s(-1)^{x_1+x_2}}\left(\phi_x-\frac{\sqrt{k}}{2}\right)+\frac{\sqrt{k}}{2}$, where $s$ is a real parameter, when $\beta\rightarrow 0$, which prohibits the mass term for $\phi_x$. The finite temperature correction contributes a parametrically small mass term $2\sqrt{\beta}(\mu-2U)\phi_x^2$ (see eq.~\eqref{eq:prop}), causing the checkerboard mode to condense when $U>\mu/2$, inducing a gas/solid phase transition. The result, furthermore, seems to be valid to all orders in $k$, as long as $\phi_x$ is assumed continuous. The latter assumption is not true for finite $k$, and it is not clear whether (possibly non-perturbative) corrections invalidate the conclusion at finite $k$.

Even if the critical $U_c=\mu/2$ receives a finite $k$ correction, we would like to point to an unusual situation where the high temperature gas phase is accompanied by a large correlation length, setting it apart from the normal gas where high temperature contributes to a small correlation length -- the thermal screening length. Study of this curious phenomena as well as finite $k$ corrections to the formula $U_c=\mu/2$ is left for future study.

\subsection{Large $k$ Quantum AIM model}

We consider $k$ bosonic flavors $b_{x,\alpha}$ on a square lattice with nearest-neighbor density interaction,
\begin{gather}
  n_{x,\alpha}=b^\dagger_{x,\alpha}b_{x,\alpha},\quad n_x=\sum_{\alpha=1}^{k}n_{x,\alpha},\quad
  H = -t\sum_{\langle i j\rangle,\alpha}\!\!\big(b^\dagger_{x,\alpha} b_{y,\alpha}+{\rm h.c.}\big)
      +\mu \sum_{x,\alpha} n_{x,\alpha}
      +\frac{U}{k}\sum_{\langle i j\rangle}\sum_{\alpha,\beta} n_{x,\alpha} n_{y,\beta}.
\end{gather}
% \TS{Is there a sign error in the $\mu$-term?}
For convenience we can rewrite the partition function of this model in terms of the path integral over bosonic variables $b_{x,\alpha},\bar{b}_{x,\alpha}$
\begin{gather}
Z=\!\int\!\mathcal{D}\bar b\,\mathcal{D}b\;
\exp\!\left[-\!\int_0^\beta\! d\tau\left(
\sum_{x,\alpha}\bar b_{x,\alpha}(\partial_\tau+\mu)b_{x,\alpha}
- t\!\sum_{\langle x,y\rangle,\alpha}\!(\bar b_{x,\alpha} b_{y,\alpha}+{\rm c.c.})
+\frac{U}{k}\!\sum_{\langle x,y\rangle,\alpha,\beta}\!\bar b_{x,\alpha} b_{x,\alpha}\bar b_{y,\beta} b_{y,\beta}
\right)\!\right]\!.
\end{gather}
We decouple the nearest-neighbor density-density interaction using the Hubbard-Stratonovich auxiliary field $\sigma_x(\tau)$,
\begin{gather}
  \exp\!\left\{-\frac{U}{k}\!\int_0^\beta\! d\tau\sum_{\langle ij\rangle} n_x n_y\right\}
  = \!\int\!\mathcal{D}\sigma\;
  \exp\!\left\{-\!\int_0^\beta\! d\tau\left[
     \frac{k}{4U}\sum_{x,y}\sigma_x \mathcal{V}^{-1}_{x,y}\sigma_y
    +\sum_x \sigma_x n_x
  \right]\!\right\},
\end{gather}
where $\mathcal{V}_{xy}=-1/2$ for the nearest neighbor $x,y$\footnote{The convergent Gaussian integral is given by rotating $\sigma_x \to i \sigma_i$, which does not affect the saddle-point solution.}.
% is the kernel that reproduces the nearest-neighbor coupling (e.g., one may take $\mathcal{V}=-\Delta$ proportional to the lattice Laplacian). 
In the large-$k$ saddle we take a time-independent, two-sublattice ansatz
\begin{gather}
  \sigma_x=\sigma+(-1)^i\eta,\qquad (-1)^i=\begin{cases}+1,& i\in A\\-1,& i\in B\end{cases}.
\end{gather}
With this static background, the quadratic boson action leads to two bands
\begin{gather}
  E_{k}^{\pm}=\mu+\sigma\pm\sqrt{\eta^2+\varepsilon_{k}^2},\qquad
  \varepsilon_{k}=-2t(\cos k_x+\cos k_y).
\end{gather}
These are bosonic single-particle bands obtained from folding into the two-sublattice Brillouin zone $(k_x,  k_y) \in [-\pi/2, \pi/2]\times[0,2\pi]$. The large-$k$ free-energy density per flavor is
\begin{gather}
  -\frac{T}{k N}\log Z = 
  -\frac{1}{8U}\big(\sigma^2-\eta^2\big)
  + T\int_{\rm BZ_{\frac12}}\frac{d^2k}{(2\pi)^2}\sum_{s=\pm}\log\!\left[1-e^{-E_{k}^{\,s}/T}\right],
\end{gather}
which yields the saddle-point equations (with $n_B(x)=\frac{1}{e^{x/T}-1}$):
\begin{gather}
  \frac{\sigma}{4U}=\int_{\rm BZ_\frac12}\frac{d^2k}{(2\pi)^2}\Big[n_B(E_{k}^{+})+n_B(E_{k}^{-})\Big], \quad 
  \frac{\eta}{4U}=-\int_{\rm BZ_\frac12}\frac{d^2k}{(2\pi)^2}\;\frac{\eta}{\sqrt{\eta^2+\varepsilon_{k}^2}}
  \Big[n_B(E_{k}^{+})-n_B(E_{k}^{-})\Big].
\end{gather}
Let us work in the high-$T$ limit, and, first of all, assume that $\eta = 0$. The first saddle-point equation yields
\begin{align}
    \sigma = 2\sqrt{U T} - \frac12(2U +\mu) + \frac{3 \sqrt{2} \mu ^2+48 \sqrt{2} t^2+7 \sqrt{2} U^2-6 \sqrt{2} \mu  U}{48 \sqrt{TU}} + \mathcal{O}(T^{-1})
\end{align}
And the last equation has a non-trivial solution with $\eta \neq 0 $ when 
\begin{gather}
  \frac{1}{2U}
  =
  -\int_{\rm BZ_\frac12}\frac{d^2k}{(2\pi)^2}\;
  \frac{\sinh\!\big(|\varepsilon_{k}|/T\big)}
  {|\varepsilon_{k}|\,[\cosh\!\big((\mu+\sigma)/T\big)\, - \cosh\!\big(|\varepsilon_{k}|/T\big)]}.
\end{gather}
Substituting $\sigma$ in the high temperature expansion we find that the previous equation at high temperatures is \begin{gather}
     \frac{1}{4U} = \frac{1}{4U} + \frac{U - \frac12 \mu}{4 \sqrt{T U}} + \mathcal{O}\left(T^{-1}\right)
\end{gather}
And plugging this  we get that $U_{\rm crit} =\frac12\mu$. Thus for $U>\frac{\mu}{2}$ we would have a solution $\eta \neq 0$ and lower free energy.

\section{Field Theory transformation}
Consider the grand canonical ensemble given by
\begin{equation}
Z=\sum \frac{z^N}{N!}\left(\prod_i\int d^D\vec x_i\right) e^{- \sum_{i<j=1}^N U(|\vec x_i-\vec x_j|)\beta}
\end{equation}
We can write it formally as a path integral
\begin{equation}
Z=\int Dn(\vec x)D\sigma(\vec x) e^{-\beta\int d^D\vec x\int d^D\vec y n(\vec x) U(|\vec x-\vec y|)n(\vec y)+\int d^D\vec x\;\left( ze^{i\sigma(\vec x)}-{i\sigma(\vec x)n(\vec x)}\right)}\;.
\end{equation}
Expanding the above exponent to order $z^N$, it is easy to see that the two forms agree order-by-order in $z$. 
If $z$ is large, we can treat the integral over $\sigma(\vec x)$ by a saddle point approximation. The saddle point equations for $\sigma(\vec x)$ imply
\begin{equation}
  z e^{i\sigma(\vec x)}=n(\vec x)  
\end{equation}
so that 
\begin{equation}
Z\approx \int Dn(\vec x) \exp\left(-\beta\iint d^D\vec x d^D\vec y\; n(\vec x) U(|\vec x-\vec y|)n(\vec y)+\int d^D\vec x\;n(\vec x)\left(1 -\log(n(\vec x)/z)\right)\right)\;.
\end{equation}
The exponent of the integrand above expression is $-\mathcal F\beta$ that we study in the main text.

\section{Monte Carlo Simulation Details}

Monte Carlo simulations of the Arithmetic Ising Model were done using the standard metropolis update of the spins $n_x$. Since spins are unsuppressed all up to a value $n_x\sim T$ with a large $T$, proposals which change $n_x$ by $\pm 1$ are not efficient when temperature $T$ is large. Therefore we used the update $n_x\rightarrow n_x\pm \Delta n$ where $\Delta n$ was drawn from a uniform distribution of integers from $1$ to $\left\lfloor \frac{T}{10}\right\rfloor$, where $T$ is the temperature. 

All the simulations were then thermalized with $2\times 10^5$ thermalization sweeps. To avoid long lasting metastable domain wall states, the simulations were initialized with an ordered checkerboard state with spins $n_x=0$ on one sublattice and $n_x=2+\lfloor T\rfloor$ on the other. $10^5$ measurments of observables were taken with $300$ decorrelation sweeps in between. The errors were estimated using the binned Jackknife method.

\end{appendix}

\end{document}